\documentclass[aps,pra,reprint,groupedaddress]{revtex4-1}

\usepackage{amsmath}
\usepackage{color}
\usepackage{amsfonts}
\usepackage{graphicx}

\newcommand{\mat}[1]{\mathbf{#1}}
\renewcommand{\vec}[1]{\boldsymbol #1}
\newcommand{\op}[1]{\hat{#1}}
\newcommand{\opd}[1]{{\hat{#1}^\dagger}}

\newcommand{\kappaMW}{\kappa_{\mu\mathrm{w}}}
\newcommand{\kappaMM}{\kappa_{\mathrm{mm}}}
\newcommand{\kappaMWint}{\kappa_{\mu\mathrm{w,int}}}
\newcommand{\kappaMMint}{\kappa_{\mathrm{mm,int}}}
\newcommand{\omegaMW}{\omega_{\mu\mathrm{w}}}
\newcommand{\omegaMM}{\omega_{\mathrm{mm}}}
\newcommand{\omegaPump}{\omega_{\mathrm{p}}}
\newcommand{\Pheating}{P_{\text{heating}}}
\newcommand{\nPump}{n_{\mathrm{p}}}
\newcommand{\Ebit}{E_{\text{qbit}}}

\newcommand{\kappaPump}{\kappa_{\mathrm{p}}}
\newcommand{\ie}{\emph{i.e.},~}

\begin{document}


\title{Millimeter-wave interconnects for microwave-frequency quantum machines}


\author{Marek Pechal}
\email[]{mpechal@stanford.edu}

\author{Amir H. Safavi-Naeini}
\email[]{safavi@stanford.edu}

\affiliation{Ginzton Laboratory, Stanford University, Stanford, California 94305, USA}


\date{\today}

\begin{abstract}
Superconducting microwave circuits form a versatile platform for storing and manipulating quantum information. A major challenge to further scalability is to find approaches for connecting these systems over long distances and at high rates. One approach is to convert the quantum state of a microwave circuit to optical photons that can be transmitted over kilometers at room temperature with little loss. Many proposals for electro-optic conversion between microwave and optics use optical driving of a weak three-wave mixing nonlinearity to convert the frequency of an excitation. Residual absorption of this optical pump leads to heating, which is problematic at cryogenic temperatures. Here we propose an alternative approach where a nonlinear superconducting circuit is driven to interconvert between microwave-frequency ($7\times10^9$ hertz) and millimeter-wave-frequency photons ($3\times 10^{11}$ hertz). To understand the potential for quantum state conversion between microwave and millimeter-wave photons, we consider the driven four-wave mixing quantum dynamics of nonlinear circuits. In contrast to the linear dynamics of the driven three-wave mixing converters, the proposed four-wave mixing converter has nonlinear decoherence channels that lead to a more complex parameter space of couplings and pump powers that we map out. We  consider physical realizations of such converter circuits by deriving theoretically the upper bound on the maximum obtainable nonlinear coupling between any two modes in a lossless circuit, and synthesizing an optimal circuit based on realistic materials that saturates this bound. Our proposed circuit  dissipates less than $10^{-9}$ times the energy of current electro-optic converters per qubit. Finally, we outline the quantum link budget for optical, microwave, and millimeter wave connections, showing that our approach is viable for realizing interconnected quantum processors for  intracity or quantum datacenter environments.
\end{abstract}


\maketitle


\section{Introduction}

 As engineered quantum systems of ever-greater complexity are realized in labs across the world, it is important to address the practical challenges facing large-scale machines. One vision of scaling involves the construction of a quantum internet~\cite{Kimble2008a} enabled by nodes connected via links that distribute entanglement at a high rate. High bandwidth quantum interconnects able to faithfully transmit quantum information may facilitate implementation of large-scale quantum systems composed in a modular way of specialized subsystems~\cite{Reiher2016}. In the superconducting circuits architecture \cite{Devoret2013a} -- one of the most promising platforms for engineered quantum systems -- short-distance connections use on-chip coplanar stripline or machined hollow waveguides. But options for connecting qubits across longer distances are limited. Losses in long superconducting waveguides and the difficulty of cooling them to millikelvin temperatures pose a challenge in distributing entanglement across a network. Estimates of losses in hollow metallic waveguides at dilution cryostat temperatures \cite{Kurpiers2016,Xiang2016,Vermersch2017} indicate that direct links longer than tens of meters are impractical. Moreover, the low electromagnetic carrier frequency used to transmit the quantum information means that these wavelength-scale waveguides are physically large and are affected by  noise due to the significant thermal photon occupation at higher temperatures.

A commonly suggested approach to circumvent the three serious limitations of microwave-frequency interconnects -- high loss, large size, and excess thermal noise -- is to convert the quantum information processed in the microwave-frequency quantum machine into optical photons for transmission. Optical fibers provide a medium that has losses below $0.2~\text{dB/km}$ over several terahertz of bandwidth with essentially no background thermal photon occupation. There is now a wide-ranging effort to realize quantum microwave-to-optical converters. Often such converters use strong optical pumping of a weak electromagnetic nonlinearity to shift signals across the five orders of magnitude disparity in energy.  
The high optical pump power introduces significant heat dissipation per converted qubit which ultimately limits transmission rates.

Here we propose and investigate the feasibility of an alternative approach, \ie conversion of the microwave signals to millimeter wave (mm-wave) frequencies in the range of hundreds of gigahertz. As we will show, an advantage in comparison with microwave connections is that links useful for quantum information transfer can be realized at much more modest cryogenic temperatures and with a smaller footprint. Compared to optical links, it avoids the need for strong optical pumping of the electro-optic converters and associated issues with heating and optical absorption in superconductors that may limit scaling. 

We start in section~\ref{sec:motivation} by describing the merits and limitations of current approaches. We are particularly interested in  conversion rate limitation due to the energy consumption of existing methods based on three-wave mixing. We present a theoretical description of mm-wave systems that use four-wave mixing to convert information between mm-wave and microwave frequencies in section~\ref{sec:mmwave_theory}. To understand how such a converter circuit can make orders of magnitude higher transmission rates possible, we place bounds on the energy consumption. We do this by deriving a general theorem in section~\ref{sec:sumrule} that sets an upper bound on the effective interaction rate achievable between any two modes in an arbitrary lossless quantum circuit incorporating nonlinear elements. In section~\ref{sec:maxcouple} we use this sum rule to arrive at estimates of circuit parameters and propose circuits that saturate this bound and enable significant interaction rates with current state-of-the-art materials. We calculate the energy dissipated per converted qubit in section~\ref{sec:mmwave_Ebit}. Note that unlike the three-wave mixing processes which become linear at large pump powers, nonlinear effects become more important with increasing pump power in conversion schemes based on four-wave mixing. This means that there is a trade-off between high conversion rates and linearity. To understand the resulting upper limits of achievable conversion rates, we need to carefully consider the nonlinear effective Hamiltonian and account for shot noise decoherence channels which result from treating the pump as a quantized mode rather than a classical drive. The more complex nonlinear dynamics leads to a rich parameter space that we explore in section~\ref{sec:mmwave_Ebit}. We note that these nonlinear effects were not considered in the analysis of four-wave mixing conversion experiments demonstrated to-date \cite{Zakka-bajjani2011,Li2015a}, as they operated in a regime where fully classical treatment of the pump was satisfactory. Finally, in sections~\ref{sec:losses} and \ref{sec:footprint} we compare the mm-wave approach to the pure microwave and optical methods by taking into account the expected end-to-end losses and the physical space required by the link.

\section{Optical-to-microwave conversion energy and motivation for mm-waves}\label{sec:motivation}

Parametric processes can be used for quantum conversion of excitations with vastly different energies~\cite{Louisell1961,Kumar1990}. For conversion between optical and microwave frequency excitations, the three-wave mixing process provided by optomechanics~\cite{Safavi-Naeini2010a,Wang2012,Andrews2014a,Balram2015b,Vainsencher2016a}, electro-optics~\cite{Tsang2010,Rueda2016,Witmer2017}, and magneto-optics~\cite{Hisatomi2016} can be used. We describe these processes using an interaction Hamiltonian $g_0 \op{a}^\dagger \op{a} (\op{b}+\op{b}^\dagger)$, where $\op{a}$ and $\op{b}$ are the annihilation operators for the optical mode and the microwave-frequency resonance. The basic requirement in all cases is that the parametrically enhanced interaction rate $g_0 \sqrt{\nPump}$, where $\nPump$ is the mean optical pump photon occupation, equals the geometric mean of the losses into the microwave and optical channels. This condition is captured by the requirement of unity cooperativity, $C\equiv 4g_0^2 \nPump / \kappa_\text{opt} \kappa_{\mu\text{w}} = 1$ (see appendix \ref{app_convEvolution}). Here $\kappa_{\mu\text{w}}$ is the loaded microwave bandwidth, which is the bandwidth of conversion. A pump photon occupation of $\nPump$ leads to a heating rate of $\Pheating=\nPump\kappa_\text{opt,i}$ where $\kappa_\text{opt,i}$ is the part of the optical linewidth due to absorption and scattering into the cryogenic environment. From here we can calculate an energy per ``bit'' or temporal mode that is converted:
\begin{equation}\label{eq:EbitElectrooptical}
\Ebit=\hbar\omega_\text{opt}\left(\frac{\kappa_\text{opt,i}}{g_{0}}\right)^{2}\frac{1}{\eta^{2}(1-\eta^{2})},
\end{equation}
where the efficiency $\eta$ is defined by the relation $\kappa_\text{opt,i}=(1-\eta^{2})\kappa_\text{opt}$. $\Ebit$ is an important figure of merit as it sets a bound on the achievable conversion rate due to the maximum allowed heat load $P_{\mathrm{max}}$ in the cryostat. This upper bound is given by $P_{\mathrm{max}}/\Ebit$, so we should aim to minimize $\Ebit$. Note that since we can in principle have a large number of converters operating in parallel, it is the energy per bit which determines the throughput rather than just the bandwidth of the converter. As the time required to convert one qubit is given by the inverse of the converter's bandwidth $\Delta$, we can estimate $\Ebit$ in terms of the power $\Pheating$ dissipated by the device operating in a continuous mode as $\Ebit = \Pheating/\Delta$. 

To outline the technical challenges in optimizing $\Ebit$, we note an interesting relation between the quantum converters and classical electro-optic modulators. In classical electro-optic modulators where an electrical signal is used to modulate an optical field, the important figure of merit is the electrical energy required to switch the optical beam. This is approximately equal to the charging energy of the capacitor  $C_\text{m}$ surrounding an electro-optic medium, $E_{\text{cbit}}\propto C_{\text{m}}V_{\pi}^{2}$~\cite{Miller2012a}. The voltage $V_\pi$ needed to switch the state is determined by the geometry and the material's nonlinear properties. We can estimate $V_\pi$ for the optically resonant systems typically considered for quantum conversion by noting that a shift in frequency on the order of the optical linewidth $ \kappa_\text{opt}$ is needed to switch the beam from an on to an off state, and so we set $V_\pi= \kappa_\text{opt}/g_\text{V}$ where $g_\text{V}$ is the shift in the optical cavity frequency per volt. This parameter is related simply to the modal coupling rate as $g_0=g_\text{V}V_\text{zp}$, where $V_\text{zp}$ is the zero point voltage fluctuation amplitude, leading to a relation 
\begin{equation}
\Ebit\propto\frac{\omega_\text{opt}}{\omega_{\mu\text{w}}}E_\text{cbit}.\label{eqn:qbit_cbit}
\end{equation}
A surprising aspect of this relation is that $\Ebit$ and $E_\text{cbit}$ correspond to different types of loss. $\Ebit$ accounts for the heating when \emph{optical} energy is dissipated in the modulator, while $E_\text{cbit}$ is the \emph{microwave} energy required to switch a classical modulator. The energy consumption of classical electro-optic modulators, $E_\text{cbit}$, has been subject to intense optimization by industry and academia in the last decade for interconnect and data center applications. Current world records are on the order of femtojoules~\cite{Miller2017}. A quantum converter leveraging the best available classical technology would therefore be limited to $\Ebit\approx 1-10~\text{nanojoules}$ due to the ``quantum conversion'' prefactor $\omega_\text{opt}/\omega_{\mu\text{w}}$ in equation~(\ref{eqn:qbit_cbit}). Similar and slightly lower numbers are obtained for recent experimental efforts utilizing mechanical resonators~\cite{Andrews2014a,Balram2015b,Vainsencher2016a} as intermediaries. From here we conclude that reducing $\Ebit$ can be accomplished either by improving the magnitude of the nonlinearity ($g_0/\kappa_\text{opt,i}$) or by reducing the carrier frequency $\omega_\text{opt}$. As we will see below, moving the carrier from optical ($10^{14}-10^{15}$ hertz) to mm-wave frequencies ($10^{11}$ hertz) achieves both.

\section{Microwave to mm-wave conversion theory}\label{sec:mmwave_theory}
The envisaged method for efficiently converting the quantum state of photons between gigahertz and  hundreds of gigahertz with small added noise is an extension of the impedance matched three-wave mixing scheme proposed in~\cite{Safavi-Naeini2010a}. Variants of this technique have been used for interconverting microwave-frequency photons~\cite{Abdo2013,Lecocq2016,Ockeloen-Korppi2016,Fink2015a}, optical-frequency photons~\cite{Dong2012a,Hill2012}, and there are on-going experiments to demonstrate conversion between optical and microwave frequencies~\cite{Vainsencher2016a,Balram2015b,Bagci2014,Andrews2014a}. In superconducting circuits, the more natural nonlinearity is a nonlinear kinetic inductance which leads to four-wave mixing. Conversion between modes by four-wave mixing was initially considered by Louisell \emph{et al.}~\cite{Louisell1961} and quantum converters at microwave~\cite{Zakka-bajjani2011} and optical frequencies~\cite{Li2015a} have been experimentally demonstrated. The key idea is that pumping at an intermediate frequency couples microwave and a mm-wave resonant modes $a$ and $b$ and induces an effective beam-splitter interaction $g(a^{\dagger}b + b^{\dagger}a)$. The two modes are coupled to open transmission lines with rates $\kappaMW$ and $\kappaMM$. A fraction of the power sent through the microwave line is up-converted by the pumped system and emitted into the mm-wave line. An input-output theory analysis (see Appendix \ref{app_convEvolution}) shows that the theoretical efficiency of this conversion process reaches unity when the cooperativity parameter $C = 4g^2/\kappaMW\kappaMM$ is equal to one. The beam-splitter coupling strength required for perfect conversion is therefore given by
\begin{equation}\label{eq_unitConvEffCoupling}
g^2 = \frac{1}{4}\kappaMW\kappaMM.
\end{equation}
\begin{figure}
\includegraphics[width=7.0cm]{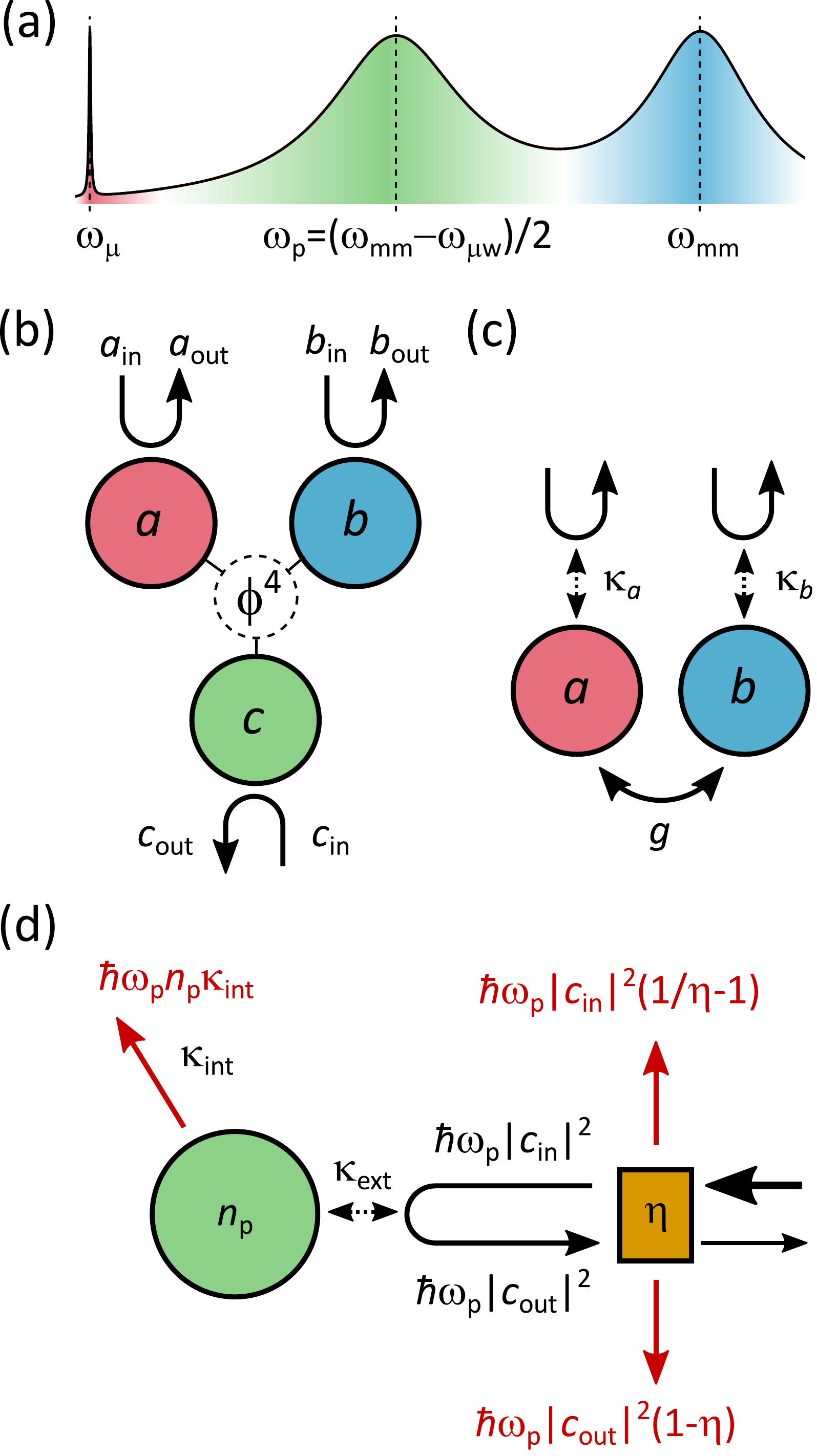}
\caption{(a) Spectrum of the system, showing the three relevant modes and (b) a diagram representing their mutual interaction via the $\phi^4$ nonlinearity of the kinetic inductor and coupling to their respective continuum input/output fields. (c) The effective coupled system of modes $a$ and $b$ when $c$ is coherently displaced by a strong pump signal. (d) Model of heat dissipation due to the pump tone at the base stage of the cryostat. The terms shown in red are, from top left to bottom right, the internal dissipation in the pump mode and the pump line losses of the incoming and reflected field.}
\label{fig_diagrams}
\end{figure}

A kinetic inductor -- typically a thin wire made of a suitable material such a niobium nitride (NbN) or niobium titanium nitride (NbTiN) -- is a weakly nonlinear element approximately described by the Hamiltonian
\[
  \op{H}_{\mathrm{KI}} = \frac{\op{\phi}^2}{2L} - \frac{\op{\phi}^4}{4L^3 I_{*}^2},
\]
where $\op{\phi}$ is the difference of the flux variable $\int \op{V}\,\mathrm{d}t$ across the inductor, $I_{*}$ is the \emph{cross-over current} \cite{Vissers2015} and $L$ the wire's low-current inductance. The term quadratic in $\op{\phi}$ corresponds to a linear inductor and we can include it into the free Hamiltonian $\op{H}_0$ of the circuit together with terms due to the other linear circuit elements. Diagonalization of $\op{H}_0$ then yields the normal modes of the circuit, two of which will be the microwave mode $a$ and the mm-wave mode $b$. The quartic term can then be treated as a perturbation inducing four-wave mixing. In particular, it enables two photons of a strong pump at a frequency $\omegaPump=(\omegaMM-\omegaMW)/2$ to combine with a microwave photon at $\omegaMW$ and produce a mm-wave photon at $\omegaMM$.

We assume that to couple the pump signal into the system, it is designed to have a third normal mode $c$ at the pump frequency $\omegaPump$, as shown schematically in Figs.~\ref{fig_diagrams}(a,b). We will write the operator $\op{\phi}$ as a combination of the ladder operators $\op{a},\op{b},\op{c}$ of the three relevant modes: $\op{\phi} = \phi_a(\op{a}+\op{a}^{\dagger}) + \phi_b(\op{b}+\op{b}^{\dagger})+\phi_c(\op{c}+\op{c}^{\dagger})$. We perform the rotating wave approximation (RWA), retaining only energy-conserving terms in $\op{\phi}^4$. After putting the result in normal order, we get terms which are second and fourth order in the ladder operators. The resulting form of the Hamiltonian is shown in full detail in appendix \ref{app_fullHamiltonian}. 

The quadratic terms describe a dressing of the normal mode frequencies by the nonlinearity and we will absorb them into $\op{H}_0$. We are then left with a Hamiltonian containing terms such as $\op{a}^{\dagger}\op{a}^{\dagger}\op{a}\op{a}$ describing anharmonicities of the modes, cross-Kerr coupling terms of the form $\op{a}^{\dagger}\op{a} \op{b}^{\dagger}\op{b}$ and an energy exchange term $g_0(\op{a}\op{b}^{\dagger}\op{c}\op{c}+\mathrm{H.c.})$ 

If the mode $c$ is driven resonantly by a coherent pump tone, this results in a constant displacement of $\op{c}$ to $\op{c}+\sqrt{\nPump}$ in the reference frame co-rotating with the pump. Here $\nPump$ is the mean number of photons induced in the mode by the pump. As in other four-wave mixing conversion schemes, we assume that $\nPump\gg 1$ and to first approximation, we only keep the largest coupling terms proportional to $\nPump$ in the Hamiltonian. In our initial analysis of the system, we neglect the lower-order terms. Later in section~\ref{sec:mmwave_Ebit} and appendix~\ref{app_measDeph}, we analyze the effects of the higher order terms in $\op{c}$, which cause nonlinear behavior of the pump mode, and the cross terms such as $\op{a}^{\dagger}\op{a}\op{c}\sqrt{\nPump}$, which induce shot noise dephasing of the converted photons. This allows us to identify operating regimes in which the linear approximation is justified.

Assuming for now that nonlinearities in the Hamiltonian can indeed be neglected, displacement of the coupling term $\op{a}\op{b}^{\dagger}\op{c}\op{c}+\mathrm{H.c.}$ yields the desired beam-splitter interaction of the form $g_0 \nPump (\op{a}^{\dagger}\op{b}+\op{b}^{\dagger}\op{a})$, where the coupling strength is
\begin{equation}\label{eq_coupling}
  g_0 = -\frac{3\phi_a\phi_b\phi_c^2}{L^3 I_{*}^2}.
\end{equation}
The system can then be effectively described by two directly coupled modes, as shown in Fig.~\ref{fig_diagrams}(c).

If the system parameters are chosen such that the dressed modes $a$ and $b$ are resonant in the reference frame co-rotating with the pump tone and the matching condition 
\begin{equation}\label{eq_unitConvEffCoupling_np}
  \frac{4g_0^2 \nPump^2}{\kappaMW\kappaMM}=1
\end{equation}
is satisfied, the incoming fields $a_{\mathrm{in}}$ and $b_{\mathrm{in}}$ resonant with the modes are perfectly converted into $b_{\mathrm{out}}$ and $a_{\mathrm{out}}$, respectively.

To further evaluate the coupling $g_0$, we need to know the product $\phi_a\phi_b\phi_c^2$ of vacuum fluctuation amplitudes of $\op{\phi}$ for the three modes. Instead of calculating it for any specific circuit configuration, we will derive a very general result about vacuum fluctuation amplitudes in an arbitrary linear circuit and use it to find the maximum achievable value of $\phi_a\phi_b\phi_c^2$. We believe this result may be very useful in the process of designing quantum circuits and we therefore show its derivation in some detail in the following section. 

\section{Sum rule for vacuum fluctuation amplitudes}\label{sec:sumrule}
The process of electrical circuit quantization is usually presented in the context of Lagrangian and Hamiltonian dynamics \cite{Devoret1997a}. Although the procedure to arrive at the quantized model is straightforward, it is not immediately obvious if there is a connection between its parameters such as the vacuum fluctuation amplitudes and classical parameters of the circuits such as the impedance matrix. The answer to this question turns out to be positive and provides a convenient shortcut from the circuit diagram to the quantum Hamiltonian. It can be derived either from considering a specific canonical form of the circuit, as for example in \cite{Nigg2012a}, or more generally as follows:

An arbitrary lossless linear circuit can be represented as a network of capacitors and inductors described by a capacitance matrix $\mat{C}$ and an inductance matrix $\mat{L}$. The impedance matrix is then given by 
\[
  \mat{Z}(\omega) = (\mathrm{i}\omega \mat{C} + (\mathrm{i}\omega \mat{L})^{-1})^{-1}.
\]
This matrix is singular for $\omega$ equal to any of the resonance frequencies $\omega_1,\omega_2,\ldots$ of the circuit. A simple algebraic manipulation shows that $\mat{Z}(\omega)$ can be written as
\[
  \mat{Z}(\omega) = \mathrm{i}\omega \mat{C}^{-1/2}
  \mat{U}(\mat{D} - \omega^2)^{-1}\mat{U}^{\mathsf{T}}
  \mat{C}^{-1/2},
\]
where $\mat{U}$ is an orthogonal transformation which brings the symmetric matrix $\mat{C}^{-1/2} \mat{L}^{-1} \mat{C}^{-1/2}$ into its diagonal form $\mat{D}=\mathrm{diag}(\omega_1^2,\omega_2^2,\ldots)$. From here it follows that the residues of the impedance matrix poles are given by
\[
  \mathrm{res}_{\omega_k}{Z_{ij}} = -\frac{\mathrm{i}}{2}
  (\mat{C}^{-1/2}\mat{U})_{ik} (\mat{C}^{-1/2}\mat{U})_{jk}.
\]
The matrix $\mat{U}$ is closely related to the canonical transformation which diagonalizes the circuit's Hamiltonian $H = \frac{1}{2}\vec{q}^{\mathsf{T}}\mat{C}^{-1}\vec{q}+\frac{1}{2}\vec{\phi}^{\mathsf{T}} \mat{L}^{-1}\vec{\phi}$ written in terms of the vectors $\vec{q}$ and $\vec{\phi}$ of node charges and fluxes. Indeed, if we define
\[
  \op{a}_k = \frac{
  \omega_k^{1/2}(\mat{C}^{1/2}\mat{U})_{ik} \op{\phi}_i + 
  \mathrm{i}\omega_k^{-1/2}(\mat{C}^{-1/2}\mat{U})_{ik} \op{q}_i
  }{\sqrt{2\hbar}},
\]
we can easily verify that $\op{H}=\sum_{k}\hbar\omega_k(\op{a}_k^{\dagger}\op{a}_k+1/2)$ and the operators $\op{a}_k$ satisfy the canonical commutation relations $[\op{a}_i,\op{a}_j^{\dagger}]=\delta_{ij}$, assuming that $[\op{\phi}_i,\op{q}_j]=\mathrm{i}\hbar\delta_{ij}$. By inverting this relation, we obtain 
\[
  \op{\phi}_i = \sum_{k}\phi_{i}^{(k)}(\op{a}_k+\op{a}_k^{\dagger}),
\]
where the vacuum fluctuation amplitudes $\phi_{i}^{(k)}$ associated with node $i$ and normal mode $k$ are given by $(\mat{C}^{-1/2}\mat{U})_{ik}\sqrt{\hbar/2\omega_k}$. Combining this result with the expression for the impedance matrix residues, we get
\begin{equation}\label{eq_vacAmpVsZ}
  \mathrm{res}_{\omega_k}{Z_{ij}} = -\frac{\mathrm{i}\omega_k}{\hbar}
  \phi_{i}^{(k)}\phi_{j}^{(k)}.
\end{equation}

This equation provides a simple way to directly access the vacuum fluctuation amplitudes of the node fluxes from the impedance of the circuit. The residues of the impedance matrix in general depend on the capacitive as well as inductive elements of the circuit. In typical superconducting circuits, the nonlinear elements are the inductors and it is therefore natural to ask if there is any fundamental relation linking the inductance matrix to the vacuum fluctuation amplitudes, independently of the capacitance matrix. To derive such a relation, we divide Eq.~(\ref{eq_vacAmpVsZ}) by $\omega_k^2$ and sum over all modes $k$. The resulting sum on the left-hand side can be evaluated using Cauchy's integral theorem. The form of $\mat{Z}(\omega)$ implies that the complex integral of $\mat{Z}(\omega)/\omega^2$ along a circle of radius $R\to\infty$ asymptotically approaches zero. This means that the sum of residues of $\mat{Z}(\omega)/\omega^2$ vanishes. These residues are $\mathrm{res}_{\pm\omega_k}\mat{Z}/\omega_k^2$ due to the poles of $\mat{Z}$ at $\pm\omega_k$ plus the additional residue at $\omega=0$ introduced by the $1/\omega^2$ term. This last residue is equal to $\mat{Z}'(0)=\mathrm{i}\mat{L}$. We therefore get
\[
  \sum_{k} \frac{\phi_{i}^{(k)}\phi_{j}^{(k)}}{\hbar\omega_k}
  = \frac{1}{2}L_{ij}.
\]

The nonlinear terms in the Hamiltonian are directly related to the vacuum fluctuation amplitudes $\Delta\phi_{ij}^{(k)}\equiv \phi_{i}^{(k)}-\phi_{j}^{(k)}$ of flux differences across the nonlinear components. For these, the last equation implies
\begin{equation}\label{eq_sumRule}
  \sum_{k} \frac{(\Delta\phi_{ij}^{(k)})^2}{\hbar\omega_k}
  = \frac{1}{2} L_{ij}^{(\mathrm{eff})},
\end{equation}
where $L_{ij}^{(\mathrm{eff})}\equiv L_{ii}+L_{jj}-L_{ij}-L_{ji}$ is the equivalent inductance we would measure between nodes $i$ and $j$.

\begin{figure}
\includegraphics[width=7.5cm]{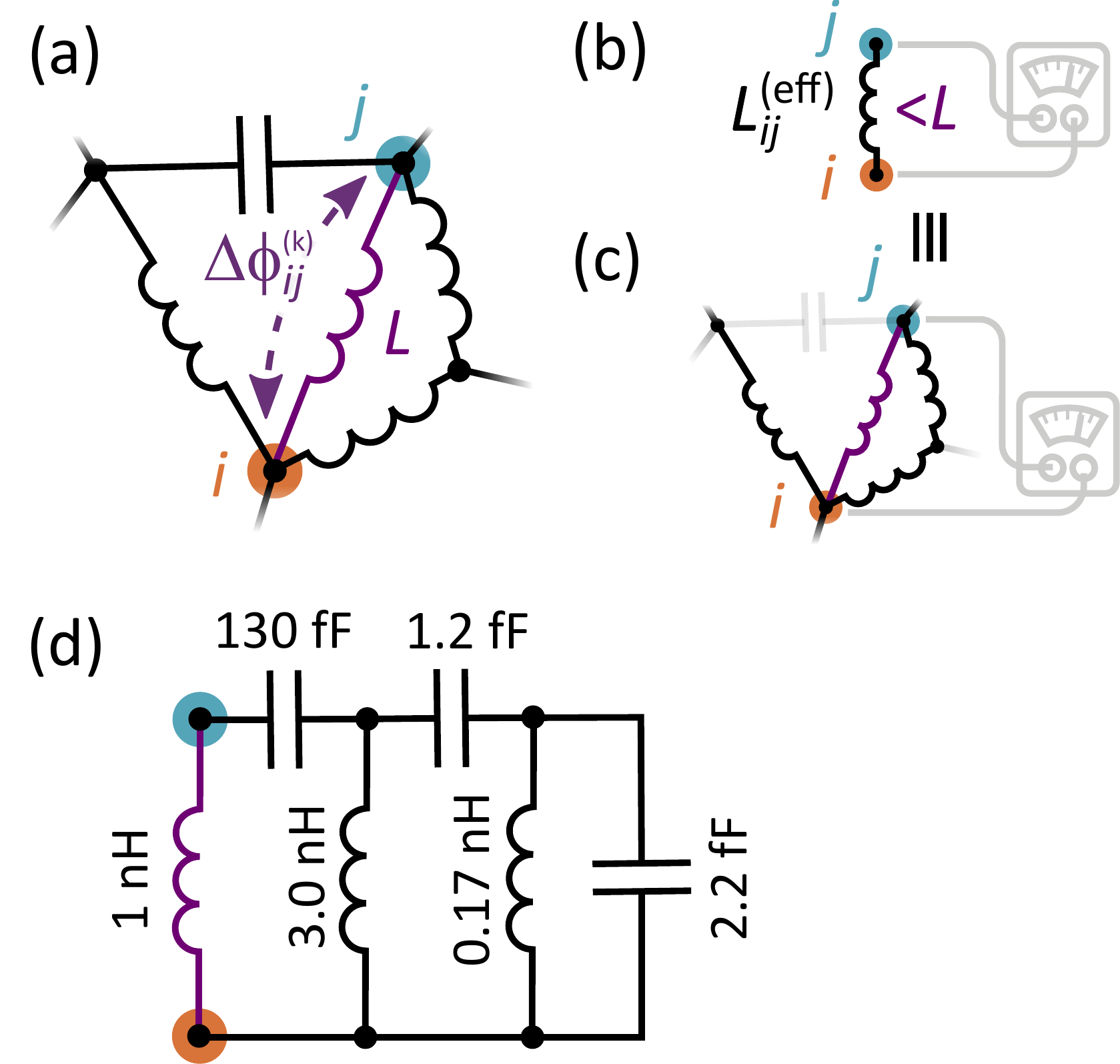}
\caption{(a) Schematic representation of an arbitrary linear circuit and the vacuum fluctuation amplitude $\Delta\phi_{ij}^{(k)}$ in a normal mode $k$ of the flux difference across an inductance $L$ (highlighted in purple) connecting its two nodes $i$ and $j$. According to the sum rule expressed by Eq.~(\ref{eq_sumRule}), this vacuum fluctuation amplitude is related to the effective inductance $L_{ij}^{(\mathrm{eff})}$ (b) we would measure by an ideal impedance meter in a circuit from which the capacitive elements have been removed (c). This inductance is in turn bounded from above by $L$. (d) An example of a circuit in a Cauer topology in which the product $\phi_a\phi_b\phi_c^2$ of vacuum fluctuation amplitudes across the inductance $L$ is maximized for given mode frequencies $\omega_a/2\pi = 7\,\mathrm{GHz}$, $\omega_b/2\pi = 300\,\mathrm{GHz}$, $\omega_c = (\omega_b-\omega_a)/2$ and for $L = 1\,\mathrm{nH}$.}
\label{fig_equivInductance}
\end{figure}

This equation relates elements of the inductance matrix to the vacuum fluctuation amplitudes of a flux difference between an arbitrary pair of nodes. In particular, if two nodes are directly connected by an inductance $L$ then $L_{ij}^{(\mathrm{eff})}\le L$ (with equality if and only if there is no path of inductors linking the two nodes except for this direct connection) and Eq.~(\ref{eq_sumRule}) therefore gives us an upper bound on a weighted sum of the squared vacuum fluctuation amplitudes over all modes. Since we are typically interested in their products rather than sums, we can use the inequality between the geometric and arithmetic mean to write
\begin{align}
  \prod_k (\Delta\phi_{ij}^{(k)})^{m_k} \le &
  \left(\frac{L}{2M}
  \right)^{M/2}
  \prod_k (\hbar\omega_k m_k)^{m_k/2}\text{, where}\nonumber\\
  M = &\ m_1+m_2+\ldots\label{eq_sumRuleIneq}
\end{align}
This inequality is the main result of this section. It puts an upper bound on products of vacuum fluctuation amplitudes across an inductor only in terms of its inductance and the frequencies of the modes. It therefore allows us to estimate coupling rates achievable with a given nonlinear element between modes at specific frequencies, with no reference to the details of the underlying circuit. The bound is sharp -- it is saturated when $L_{ij}^{(\mathrm{eff})}=L$ and $(\Delta\phi_{ij}^{(k)})^2/\hbar\omega_k m_k$ is identical for all $k$, that is if
\[
  \frac{(\Delta\phi_{ij}^{(k)})^2}{\hbar\omega_k} = \frac{m_k L}{2M}.
\]
Using Eq.~(\ref{eq_vacAmpVsZ}), we can rewrite this condition in terms of the impedance $Z_{ij}^{(\mathrm{eff})}$ measured between nodes $i$ and $j$:
\begin{equation}\label{eq_sumIneqSaturationCondZ}
  \mathrm{res}_{\omega_k}{Z_{ij}^{(\mathrm{eff})}} = -
  \frac{\mathrm{i}\omega_k^2 m_k L}{2M}.
\end{equation}

Circuits which saturate inequality (\ref{eq_sumRuleIneq}) can be implemented with passive circuits. This follows from the classical realizability condition in network synthesis theory \cite{Brune1931,Cauer2000} which states that a function $Z(\omega)$ can be realized as an impedance of a passive circuit if and only if $Z(-\mathrm{i}s)$ is a positive real function of $s$. The condition given by Eq.~(\ref{eq_sumIneqSaturationCondZ}) prescribes the values $r_k$ of the impedance residues at its positive real poles $\omega_k$. A function satisfying it can be written as $Z_{ij}^{(\mathrm{eff})}(\omega)=\sum_k r_k/(\omega-\omega_k)-r_k^{*}/(\omega+\omega_k)$. Since the residues $r_k$ are purely imaginary with a negative imaginary part, $Z_{ij}^{(\mathrm{eff})}(-\mathrm{i}s)$ is a positive real function. 

The requirement that $L_{ij}^{(\mathrm{eff})}=L$, in other words that there is no path of inductors shunting $L$, can be satisfied by choosing a suitable topology for the circuit -- for example Foster's second form or a Cauer topology, as shown in Fig.~\ref{fig_equivInductance}(d).

\section{Maximum coupling between the modes}\label{sec:maxcouple}

We can use the results of section \ref{sec:sumrule} to evaluate the maximum achievable beam-splitter coupling in our converter system. The quantity of interest is $\phi_a\phi_b\phi_c^2$, where $\phi_a$, $\phi_b$ and $\phi_c$ are the vacuum fluctuation amplitudes of the microwave, mm-wave and pump mode, which become $\Delta\phi_{ij}^{(a)}$, $\Delta\phi_{ij}^{(b)}$ and $\Delta\phi_{ij}^{(c)}$ in Eq.~(\ref{eq_sumRuleIneq}). In this particular case, we have $m_a=m_b=1$ and $m_c=2$ which gives us
\begin{equation}\label{eq_sumRuleIneq_satCond}
  \phi_a\phi_b\phi_c^2 \le \left(\frac{L}{8}\right)^2
  \sqrt{\hbar\omega_a}\sqrt{\hbar\omega_b}
  (2\hbar\omega_c).
\end{equation}
The maximum value of $\phi_a\phi_b\phi_c^2$ given by the right-hand side is reached if $a$, $b$ and $c$ are the only modes coupling to $L$, there is no inductive path shunting the kinetic inductor and if $\phi_a^2/\hbar\omega_a = \phi_b^2/\hbar\omega_b = \phi_c^2/2\hbar\omega_c$.

We then find that the maximum achievable exchange coupling strength $g_0$ given by Eq.~(\ref{eq_coupling}) is
\[
  g_0 = -\frac{3\hbar(\omegaMM-\omegaMW)\sqrt{\omegaMW\omegaMM}}{64 L I_{*}^2}.
\]
The other important parameters of this system are the Kerr couplings of mode $c$ -- the coefficients of terms such as $\op{c}^{\dagger}\op{c}^{\dagger}\op{c}\op{c}$, $\op{a}^{\dagger}\op{a} \op{c}^{\dagger}\op{c}$,\ldots, as defined in appendix \ref{app_fullHamiltonian}. In the configuration which optimizes the vacuum amplitude product $\phi_a\phi_b\phi_c^2$, these are related to $g_0$ by
\begin{align}
  \chi_{c} =& g_0(\omegaMM-\omegaMW)/\sqrt{\omegaMM\omegaMW}
  \label{eq_chic_vs_g0}\\
  \chi_{ac} =& 2g_0\sqrt{\omegaMW/\omegaMM}\\
  \chi_{bc} =& 2g_0\sqrt{\omegaMM/\omegaMW}
\end{align}

To estimate their numerical values, we consider a $5\,\mathrm{nm}$ thick, $50\,\mathrm{nm}$ wide and $2\,\mu\mathrm{m}$ long kinetic inductor with $L=1\,\mathrm{nH}$ \cite{Miki2009}. Assuming the same crossover current density as determined in \cite{Vissers2015} for a $20\,\mathrm{nm}$ thick and $2.5\,\mu\mathrm{m}$ wide wire, our inductor should have $I_{*}=0.05\,\mathrm{mA}$. Here and in the rest of the text, we will assume a microwave frequency of $\omegaMW/2\pi = 7\,\mathrm{GHz}$ and a mm-wave frequency $\omegaMM/2\pi = 300\,\mathrm{GHz}$. For these values, we get
\begin{align*}
  g_0/2\pi =& -170\,\mathrm{kHz},\\
  \chi_{c}/2\pi = & -1.1\,\mathrm{MHz},\\
  \chi_{ac}/2\pi = & -51\,\mathrm{kHz},\\
  \chi_{bc}/2\pi = & -2.2\,\mathrm{MHz}.
\end{align*}

{The corresponding circuit, which is characterized by the given resonance frequencies and the associated impedance residues expressed by Eq.~(\ref{eq_sumIneqSaturationCondZ}), can be built up from lumped elements using standard circuit synthesis techniques. An example implementation in a Cauer topology for the values stated above is shown in Fig.~\ref{fig_equivInductance}(d).}

\section{Energy dissipated per converted qubit and dephasing effects}\label{sec:mmwave_Ebit}

Heating generation in the converter is caused predominantly by the strong pump signal whose power in turn depends linearly on the number of pump photons $\nPump = \sqrt{\kappaMW\kappaMM}/2g_0$ required to reach unit cooperativity. To estimate the dissipated power, we use a simplified model of the pump line shown in Fig.~\ref{fig_diagrams}(d). We assume the section of the line in thermal contact with the base stage of the cryostat is characterized by a transmittivity $\eta$. This will most likely be limited by losses in the line and the elements connecting it to the superconducting circuit. Note that unlike for standard microwave drive lines, strong attenuation at base temperature to thermalize the field is not necessary here. The pump tone frequency is in the 100 GHz range and the noise in the field is therefore close to quantum-limited even if it is thermalized only at 4 kelvin, where far greater cooling power is available.

If the incoming and reflected pump photon flux at the sample are $|c_{\mathrm{in}}|^2$ and $|c_{\mathrm{out}}|^2$ and the intrinsic loss rate of the mode is $\kappa_{\mathrm{int}}$, the total power dissipated at the base stage can be expressed as a sum of the three terms highlighted in red in Fig.~\ref{fig_diagrams}(d). The photon fluxes are related to $\nPump$ by $\nPump = \kappa_{\mathrm{ext}}|c_{\mathrm{in}}|^2/(\delta^2 + \kappaPump^2/4)$ and $|c_{\mathrm{out}}|^2 = |c_{\mathrm{in}}|^2 - \nPump\kappa_{\mathrm{int}}$. Here $\kappa_{\mathrm{ext}}$ is the external coupling rate of the mode which we can control by design, $\kappaPump = \kappa_{\mathrm{ext}}+\kappa_{\mathrm{int}}$ its total linewidth and $\delta$ the detuning of the pump from the mode's resonance frequency. The dissipated power therefore depends on the loss rates as
\begin{equation}\label{eq_dissPower}
  \Pheating  = \hbar\omegaPump \nPump \left(
  \kappa_{\mathrm{int}}\eta + \frac{\delta^2 + \kappaPump^2/4}{\kappa_{\mathrm{ext}}}(1/\eta-\eta) 
  \right).
\end{equation}

We can now see that since the bandwidth $\Delta$ does not depend on the parameters of the pump mode, we should choose $\delta=0$ and $\kappa_{\mathrm{ext}}=\kappa_{\mathrm{int}}$ to minimize $\Ebit=\Pheating/\Delta$. In this case of matched internal and external coupling rates, the pump signal is fully absorbed by mode $c$ and we get $\Pheating=\hbar\omegaPump \nPump \kappa_{\mathrm{int}}/\eta$. There are, however, several other factors that constrain $\kappaPump$ and $\delta$ and we need to determine if we can in fact let $\kappa_{\mathrm{ext}}=\kappa_{\mathrm{int}}$ under these constraints.

First of all we would like to work in the regime where mode $c$ behaves as a harmonic oscillator. If the nonlinear frequency shift $\chi_c \nPump$ due to the self-Kerr term $\op{c}^{\dagger}\op{c}^{\dagger} \op{c} \op{c}$ becomes large enough to significantly change the response of the mode, it can no longer be treated as an approximately linear system. Then the approximation we made by replacing $\op{c}$ with $\sqrt{\nPump}$ in the $\op{a}\op{b}^{\dagger}\op{c}\op{c}$ coupling term may not be justified. 

A calculation of the perturbation expansion of the resonator's steady state $\rho$ in the Kerr nonlinearity $\chi_c$, as outlined in appendix \ref{app_NLcorrections}, shows that to lowest order in $\chi_c$, the state's overlap with the closest coherent state $|\alpha\rangle$ is reduced by an amount on the order of $\nPump^2\chi_c^2/(4\delta^2+\kappaPump^2)$. We will therefore require that $4\delta^2+\kappaPump^2 \ge F \nPump^2\chi_c^2$, where $F$ is a dimensionless factor quantifying the linearity of the pump mode. For linear operation, we will assume $F\gg 1$.

After substituting for $\chi_c$ from Eq.~(\ref{eq_chic_vs_g0}) and for $\nPump$ from the unit cooperativity requirement (\ref{eq_unitConvEffCoupling_np}), the linearity condition can be written as
\begin{equation}\label{eq_linConstraint}
  4\delta^2 + \kappaPump^2 \ge 
  F r\kappaMW\kappaMM,
\end{equation}
where $r$ denotes the ratio $(\omegaMM-\omegaMW)^2/4\omegaMM\omegaMW$ (for $\omegaMM/2\pi=300\,\mathrm{GHz}$ and $\omegaMW/2\pi=7\,\mathrm{GHz}$ we have approximately $r\approx 10$). This inequality presents a lower limit for $\delta$ and $\kappaPump$. If the expression on the right-hand side is smaller than $4\kappa_{\mathrm{int}}^2$ then we can set $\delta=0$ and $\kappa_{\mathrm{ext}}=\kappa_{\mathrm{int}}$ to reach the global minimum of the dissipated power $\Pheating$ in Eq.~(\ref{eq_dissPower}). Otherwise, the constrained minimum of $\Pheating$ is reached for $\delta=0$ and $\kappaPump$ which saturates Eq.~(\ref{eq_linConstraint}). We can then write the energy per bit as

\begin{widetext}
\begin{equation}\label{eq_EbitVsKappas}
  \Ebit =
  \frac{\hbar\omegaPump\sqrt{\kappaMW\kappaMM}}{2|g_0|\Delta}\times
  \left\{
  \begin{array}{ll}\displaystyle
  \kappa_{\mathrm{int}}/\eta & 
  \text{ for }Fr\kappaMW\kappaMM<4\kappa_{\mathrm{int}}^2\\
  \displaystyle
  \kappa_{\mathrm{int}}\eta +
  \frac{Fr\kappaMW\kappaMM(1/\eta-\eta)/4}{
  \sqrt{Fr\kappaMW\kappaMM}-\kappa_{\mathrm{int}}} & 
  \text{ for }Fr\kappaMW\kappaMM\ge 4\kappa_{\mathrm{int}}^2
  \end{array}
  \right.
\end{equation}
\end{widetext}

\begin{figure}
\includegraphics[width=8.5cm]{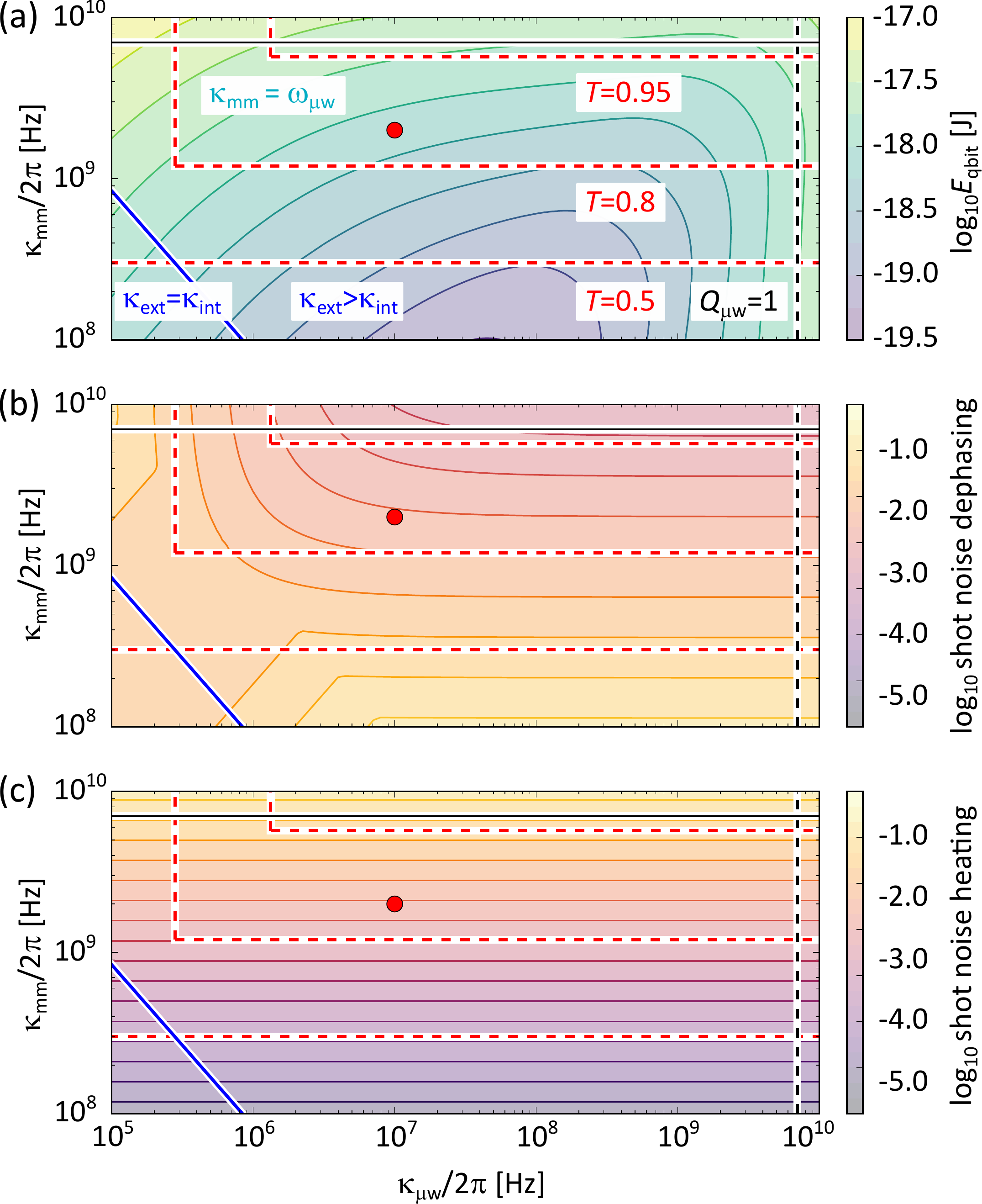}
\caption{(a) Plot of the dissipated energy per qubit $\Ebit$ as a function of the linewidths of modes $a$ and $b$, optimized with respect to the linewidth of mode $c$, as given by Eq.~(\ref{eq_EbitVsKappas}). The dashed red lines represent different values of the converter efficiency limited by the ratio between the external coupling of the modes and their internal losses. Close to the dashed black line, corresponding to unity quality factor of the microwave mode, the discrete mode model of the system becomes invalid. The solid black line indicates a value of the mm-wave linewidth equal to the microwave frequency, implying breakdown of the rotating wave approximation. The solid blue line shows the boundary between the two regimes in Eq. (\ref{eq_EbitVsKappas}). Below the line, the nonlinearity of mode $c$ is low enough to let us set $\kappa_{\mathrm{ext}}=\kappa_{\mathrm{int}}$ while above the line, Eq.~(\ref{eq_linConstraint}) forces $\kappa_{\mathrm{ext}}>\kappa_{\mathrm{int}}$. (b,c) Loss of photon state purity due to (b) dephasing and (c) heating by the pump shot noise. Evaluated using Eq.~(\ref{eq_measInducedDeph}) and Eq.~(\ref{eq_heatingPurityLoss}), respectively, with the linewidth of mode $c$ chosen to maximize $\Ebit$, as in (a). The red dot indicates the operating point chosen in our example as a compromise between the dephasing and heating effects.}
\label{fig_Ebit_MeasDeph}
\end{figure}

We  elucidate this result with the help of Fig.~\ref{fig_Ebit_MeasDeph}(a) which shows a plot of $\Ebit$ as a function of the two linewidths $\kappaMW$ and $\kappaMM$ based on Eq.~(\ref{eq_EbitVsKappas}). In this example, we again assume the mode frequencies to be $\omegaMW/2\pi=7\,\mathrm{GHz}$, $\omegaMM/2\pi=300\,\mathrm{GHz}$ and $\omegaPump=(\omegaMM-\omegaMW)/2$. For the transmittance $\eta$ of the pump line at the cryostat's base temperature, we choose $\eta=0.9$ and for the linearity parameter $F=100$. To estimate the intrinsic losses $\kappa_{\mathrm{int}}$ of mode $c$, we assume an internal quality factor $Q_{\mathrm{mm}}\approx 1000$, consistent with values previously observed in measurements of NbTiN mm-wave resonators by Endo \textit{et al.}~\cite{Endo2013}. This gives us a value $\kappa_{\mathrm{int}}/2\pi = 150\,\mathrm{MHz}$ to use in Eq.~(\ref{eq_EbitVsKappas}).

For combinations of linewidths below the blue diagonal line in Fig.~\ref{fig_Ebit_MeasDeph}(a), the required number of pump photons $\nPump\propto\sqrt{\kappaMW\kappaMM}$ is small and the resulting frequency shift of mode $c$ is small compared with its intrinsic linewidth. We can therefore choose for the pump mode  $\kappa_{\mathrm{ext}}=\kappa_{\mathrm{int}}$ to minimize $\Ebit$ with respect to $\kappa_{\mathrm{ext}}$. In this case, $\Ebit$ depends on the two linewidths $\kappaMW$ and $\kappaMM$ through $\nPump/\Delta\propto \sqrt{\kappaMW\kappaMM}/\Delta$, as expressed by the first line in Eq.~(\ref{eq_EbitVsKappas}). The lines of constant $\Ebit$ in this regime occur for  $\kappaMW\kappaMM=\text{const}$. From the general expression for the bandwidth $\Delta$ derived in Appendix \ref{app_convEvolution}, we see that $\Ebit$ depends only on the ratio of the two linewidths and reaches its minimum for $\kappaMW=\kappaMM$.

Increasing the linewidth to values above the blue line forces us to increase the linewidth of mode $c$ above the ideal value $\kappaPump=2\kappa_{\mathrm{int}}$ to keep its response approximately linear. This makes the mode overcoupled and further increases the pump power required to reach the necessary number of pump photons. The amount of power dissipated in the pump input line then increases, changing the scaling of $\Pheating$ with the linewidths of modes $a$ and $b$ from $\sqrt{\kappaMW\kappaMM}$ to roughly $\kappaMW\kappaMM$. In the regime where either $\kappaMW\ll\kappaMM$ or $\kappaMM\ll\kappaMW$, the bandwidth $\Delta$ becomes simply proportional to the lower of the two linewidths and therefore $\Ebit$ scales linearly with the larger linewidth. This explains why deep in the upper region of Fig.~\ref{fig_Ebit_MeasDeph}(a) the lines of constant $\Ebit$ become roughly parallel with the horizontal axis.

Thus, in the absence of any other constraints, we would minimize $\Ebit$ by choosing $\kappaMW=\kappaMM$ while the actual values of the linewidths do not matter as long as they satisfy $Fr\kappaMW\kappaMM<4\kappa_{\mathrm{int}}^2$. In practice, the two linewidths need to be significantly larger than the intrinsic losses of modes $a$ and $b$ to achieve a conversion efficiency $T$ close to unity. This is shown in Fig.~\ref{fig_Ebit_MeasDeph}(a) by the red dashed lines of constant efficiency which we approximate by 
\begin{equation}\label{eq_convEff}
  T\approx 1 - \max(\kappaMWint/\kappaMW,\kappaMMint/\kappaMM).
\end{equation}
To estimate the intrinsic losses $\kappaMWint$ and $\kappaMMint$, we again assume a mm-wave unloaded quality factor of $1000$ and a microwave quality factor of $10^5$. 
At the same time, both $\kappaMW$ and $\kappaMM$ must be small compared with the respective mode frequencies (quality factors need to be much higher than 1), otherwise our analysis of the converter as a system with three discrete modes is not valid. In fact, as explained below, the mm-wave linewidth needs to be small even in comparison to the microwave frequency to justify our use of the rotating wave approximation when deriving the effective Hamiltonian of the converter. 
The contour of $Q_{\mu\mathrm{m}}=1$ is shown in the plot by the black dashed line and the contour $\kappaMM=\omegaMW$ by the solid black line.

In addition to the reduction in efficiency due to intrinsic dissipation of the modes captured by the parameter $T$, we must also take into account decoherence from quantum noise processes arising from strongly pumping a fourth-order nonlinearity. We group these into type (1) and type (2) processes in accordance to whether the terms giving rise to them in the Hamiltonian scale with pump intensity as $O(\sqrt{\nPump})$ or $O(\nPump)$, respectively. We treat these two types of decoherence independently. Type (1) processes arise from terms such as $\op{c}\opd{a}\op{a}$ and $\op{c}\op{a}\opd{b}$ and are energy conserving. The effect of these terms can be interpreted as the shot noise of the pump signal causing fluctuations in the phase of the converted photons, or equivalently as measurement-induced dephasing due to the pump getting entangled with the converter modes $a$ and $b$. A perturbation theory analysis of this process is carried out in appendix \ref{app_measDeph}. The calculated loss of the photon state purity as a function of the microwave and mm-wave linewidth is plotted in Fig.~\ref{fig_Ebit_MeasDeph}(b). Type (2) processes are due to terms like $\opd{a}\opd{b}$ that are energy non-conserving by $\omegaMW$ but can lead to spontaneous emission of correlated photons when $\kappaMM$ becomes comparable to $\omegaMW$ leading to a slight violation of RWA. This type of noise is analogous to the quantum limits found in three-wave mixing nonlinear converters and the quantum back-action limit to sideband cooling~\cite{Chu1985,Marquardt2007,Wilson-Rae2007,Peterson2016}. The calculated loss of purity for the type (2) process is plotted in Fig.~\ref{fig_Ebit_MeasDeph}(c). 

Taking into account these additional constraints, we choose a typical  operating point of a converter that compromises between the effects of the two main decoherence channels shown in Figs.~\ref{fig_Ebit_MeasDeph}(b,c). At this point which is indicated in the plots by the red dot, we have $\kappaMW/2\pi=10\,\mathrm{MHz}$ and $\kappaMM/2\pi=2\,\mathrm{GHz}$. The conversion efficiency approaches $85\,\%$ with sub-attojoule dissipated energy per converted qubit. This is more than nine orders of magnitude less energy dissipation than current microwave-optical converters. These values are summarized in Table \ref{tab_ExampleOperatingPointParameters} outlining the parameters of the example operating point.

\begin{table}
\begin{tabular}{lrr|c}
quantity & & value & cf.\\
\hline
microwave frequency & $\omegaMW/2\pi$ & $7\,\mathrm{GHz}$ &\\
microwave linewidth & $\kappaMW/2\pi$ & $10\,\mathrm{MHz}$ &\\
mm-wave frequency & $\omegaMM/2\pi$ & $300\,\mathrm{GHz}$ &\\
mm-wave linewidth & $\kappaMM/2\pi$ & $2\,\mathrm{GHz}$ &\\
pump mode frequency & $\omegaPump/2\pi$ & $146.5\,\mathrm{GHz}$ &\\
pump mode linearity & $F\equiv (\kappaPump/\nPump\chi_c)^2$ & $100$ &\\
pump line transmittivity & $\eta$ & $0.9$ &\vspace{5mm}\\
\multicolumn{4}{l}{{\bf derived quantities}}\\
conversion bandwidth & $\Delta/2\pi$ & $20\,\mathrm{MHz}$ & (\ref{eq_conversionBW})\\
pump mode linewidth & $\kappaPump/2\pi$ & $4.5\,\mathrm{GHz}$ & (\ref{eq_linConstraint})\\
number of pump photons & $\nPump$ & $420$ & (\ref{eq_unitConvEffCoupling_np})\\
coupling strength & $g_0 \nPump/2\pi$ & $70\,\mathrm{MHz}$ &\\
pump mode Kerr shift & $\nPump\chi_c/2\pi$ & $-450\,\mathrm{MHz}$ &\\
dissipated power & $\Pheating$ & $100\,\mathrm{pW}$ & (\ref{eq_dissPower})\\
\hspace{10mm}in mode $c$ & & $37\,\mathrm{pW}$ & (\ref{eq_dissPower})\\
\hspace{10mm}in pump line & & $63\,\mathrm{pW}$ & (\ref{eq_dissPower})\\
energy per qubit & $\Ebit$ & $0.8\,\mathrm{aJ}$ & (\ref{eq_EbitVsKappas})\\
conversion efficiency & $T$ & $0.843$ & (\ref{eq_convEff})\\
shot noise dephasing & $\delta\mathrm{Tr}\,\op{\rho}^2$ & $0.0020$ & (\ref{eq_measInducedDeph})\\
shot noise heating & $\delta\mathrm{Tr}\,\op{\rho}^2$ & $0.0051$ & (\ref{eq_heatingPurityLoss})\\
\end{tabular}
\caption{Summary of the relevant device parameters at the example operating point (red dot in Figure~\ref{fig_Ebit_MeasDeph}).}
\label{tab_ExampleOperatingPointParameters}
\end{table}

\section{Losses and added noise in the quantum link}\label{sec:losses}

A lossy link with a transmittance $t$ at a temperature $T$ can be modeled as a beam-splitter with the same transmittance whose second input port is connected to a thermal bath. It therefore adds $N = (1-t)\overline{n}$ thermal photons to each mode of the transmitted signal, where $\overline{n}=1/(\exp(\hbar\omega/k_B T)-1)$ is the Bose-Einstein thermal photon occupation per mode at the signal's frequency $\omega$. The number of noise photons added by the link is an important figure of merit since quantum information can be only transmitted without significant coherence loss if $N\ll 1$. The transmittance $t$ itself is a less critical quantity because even with low $t$, quantum information can be sent coherently using heralding techniques at the cost of effectively reduced transmission rate.

The transmittance of the link varies exponentially with its length $l$ as $t=\exp(-\alpha l)$, where $\alpha$ is the attenuation constant. For a given photon frequency $\omega$, link temperature $T$ and attenuation constant $\alpha$, we can calculate the length of the link at which the number of added noise photons $(1-t)\overline{n}$ reaches a given threshold. The calculated lengths for microwave, mm-wave and optical links at a few selected temperatures are shown in Table~\ref{tab_addedNoise}.

At dilution cryostat temperatures, all three links can be cooled very close to their quantum ground state and the length of the link is therefore not limited by added thermal noise. At $4\,\mathrm{K}$, a temperature which can be reached with significantly less resources, the microwave link becomes unusable because a length of mere $40\,\mathrm{cm}$ introduces $0.1$ thermal photon. On the other hand, a mm-wave link at $300\,\mathrm{GHz}$ remains useful -- the added noise is below $0.01$ photon for distances up to about $20\,\mathrm{m}$ and never exceeds $0.1$ photon since the thermal occupation per mode at this temperature and frequency is only $0.03$. At $70\,\mathrm{K}$, the distance reachable with less than $0.1$ added noise photon in a $300\,\mathrm{GHz}$ link drops to $70\,\mathrm{cm}$. An optical link does not suffer from added thermal noise at all because the equilibrium thermal occupation per mode is extremely low all the way up to room temperature.

This comparison shows that the mm-wave link presents an appealing alternative to direct microwave links operable at liquid helium temperatures. The significant loss in the link means that the expected energy cost per successfully transmitted qubit is scaled up from the value given by Eq.~(\ref{eq_EbitVsKappas}) by the inverse of the link transmittance $t$. However, since the conversion energy cost is much lower than for electro-optic converters, the theoretical energy-limited transmission rate of a mm-wave link at $4\,\mathrm{K}$ still exceeds that of an optical link for distances up to several hundred meters.

\begin{table}
\begin{tabular}{|l|l||l|l|l|l|}
\cline{3-6}
\multicolumn{2}{l|}{} & 
$20\,\mathrm{mK}$ & $4\,\mathrm{K}$ & $70\,\mathrm{K}$ & $300\,\mathrm{K}$\\
\hline
$7\,\mathrm{GHz}$ & $\overline{n}$ & 
$5\times 10^{-8}$ & 11 & 210 & 890\\
(microwaves) & $\alpha\ [\mathrm{dB}/\mathrm{m}]$ & 
0.01 & 0.1 & $\gtrsim 0.1$ & $\gtrsim 0.2$ \\
& $l_{0.01}\ [\mathrm{m}]$ & 
$\infty$ & 0.04 & $\lesssim 0.002$ & $\lesssim 0.0002$ \\
& $l_{0.1}\ [\mathrm{m}]$ & 
$\infty$ & 0.4 & $\lesssim 0.02$ & $\lesssim 0.002$ \\
\hline\hline
$300\,\mathrm{GHz}$ & $\overline{n}$ & 
$\approx 0$ & 0.03 & 4.4 & 20\\
(mm waves) & $\alpha\ [\mathrm{dB}/\mathrm{m}]$ & 
0.08 & 0.08 & 0.15 & 3.8\\
& $l_{0.01}\ [\mathrm{m}]$ & 
$\infty$ & 22 & 0.07 & 0.0006 \\
& $l_{0.1}\ [\mathrm{m}]$ & 
$\infty$ & $\infty$ & 0.7 & 0.006 \\
\hline\hline
$200\,\mathrm{THz}$ & $\overline{n}$ & 
$\approx 0$ & $\approx 0$ & $\approx 0$ & $\approx 0$\\
(optics) & $\alpha\ [\mathrm{dB}/\mathrm{m}]$ & 
$0.0005$ & $0.0005$ & $0.0005$ & $0.0005$\\
& $l_{0.01}\ [\mathrm{m}]$ & 
$\infty$ & $\infty$ & $\infty$ & $\infty$ \\
& $l_{0.1}\ [\mathrm{m}]$ & 
$\infty$ & $\infty$ & $\infty$ & $\infty$ \\
\hline
\end{tabular}
\caption{Thermal photon occupations $\overline{n}$, expected link losses per meter $\alpha$ and link lengths $l_{0.01}$, $l_{0.1}$ for which the number of added thermal photons reaches 0.01 and 0.1, respectively, at microwave, mm-wave and optical frequencies at a range of selected temperatures. The $\approx 0$ thermal occupation entries stand for values lower than $10^{-10}$. The microwave waveguide losses are taken from \cite{Kurpiers2016}, the mm wave losses estimated assuming a PTFE waveguide with a loss tangent of $10^{-4}$ at room temperature \cite{Lamb1996}, $2\times 10^{-6}$ below $20\,\mathrm{K}$ and $4\times 10^{-6}$ at $70\,\mathrm{K}$ \cite{Jacob2002}. The latter source only provides loss tangent values up to $20\,\mathrm{GHz}$ at cryogenic temperatures. We assume these to be good estimates for losses up to $300\,\mathrm{GHz}$ since other sources \cite{Bur1985,Balanis1969} do not show a significant variation of the PTFE loss tangent over this frequency range at room temperature.}
\label{tab_addedNoise}
\end{table}

\section{Footprint of the quantum link}\label{sec:footprint}
Another aspect in which mm-wave quantum links can be advantageous when compared with direct microwave connections is the size of the waveguide. While WR-90 microwave waveguides operating around 10 gigahertz are relatively large with a cross section of approximately $1\,\mathrm{cm}\times 2\,\mathrm{cm}$, the dimensions of dielectric waveguides at mm-wave frequencies \cite{Dolatsha2015}  can be smaller by a factor of more than 10 determined by the frequency ratio and the dielectric constant of the used material. This means that the number of quantum channels which can be established in a given cross-section is more than hundred times higher in the mm-wave domain. By the same token, a given number of channels requires a smaller space and thermal mass which makes cooling the link to cryogenic temperatures easier.

\section{Conclusions}\label{sec:conclude}
We have theoretically analyzed the potential of quantum interconnects at mm-wave frequencies in the range of several hundred gigahertz. In contrast with direct microwave connections, mm-wave quantum links using dielectric waveguides could be operated at or even above liquid helium temperature. We have studied one particular strategy for conversion from microwave to mm-wave frequencies using four-wave mixing by a kinetic inductance nonlinearity. We derived a general upper bound for couplings between modes achievable in electrical circuits with weakly nonlinear elements and found that if the coupling in our proposed converter is optimized with respect to this bound, the conversion process is expected to require less than one attojoule per converted qubit. This is about eight orders of magnitude lower than conversion to optical photons and therefore much more suitable for operation in dilution cryostats.

\begin{acknowledgments}
The authors would like to thank Amin Arbabian, Nemat Dolatsha, Michel Devoret, for insightful conversations. M.P. is supported by the Swiss National Science Foundation. A.-H.S.N. is supported by Stanford University, a SystemX seed grant on quantum technologies as well as the Terman and Hellman fellowships.
\end{acknowledgments}

\appendix

\section{Ideal operation of the converter}\label{app_convEvolution}
Here we derive the unitary evolution of the two converter modes $a$ and $b$ and their associated input/output fields within the single-photon subspace. We neglect all coupling to the pump mode $c$ except for the effective exchange term $\op{a}^{\dagger}\op{b}+\op{a}\op{b}^{\dagger}$ induced by its displacement. The Hamiltonian we consider is
\begin{align*}
  \op{H} =& g(\op{a}^{\dagger}\op{b} + \op{b}^{\dagger}\op{a})\\
  & +
  \sqrt{\kappa_a} (\op{a}^{\dagger}\op{a}_f(t)+\op{a} \op{a}_f^{\dagger}(t)) +
  \sqrt{\kappa_b} (\op{b}^{\dagger}\op{b}_f(t)+\op{b} \op{b}_f^{\dagger}(t))
\end{align*}
The energy terms $\op{a}^{\dagger}\op{a}$ and $\op{b}^{\dagger}\op{b}$ are absent since we assume the modes are resonant in the rotating frame of the pump. The operators $\op{a}_f(t)$ and $\op{b}_f(t)$ represent the continuum of field modes coupled to $\op{a}$ and $\op{b}$, respectively.

In the single-photon subspace, we can write the state of the system in general as
\[
  |\Psi\rangle = \left(
  u_a \op{a}^{\dagger} + u_b \op{b}^{\dagger} + 
  \int \alpha_{\tau} \op{a}_f^{\dagger}(\tau) + 
  \beta_{\tau} \op{b}_f^{\dagger}(\tau) \,\mathrm{d}\tau
  \right)|0\rangle,
\]
where $\alpha_{\tau}$ and $\beta_{\tau}$ are functions describing the shape of the photon wave packet in the input/output fields. The Schr\"{o}dinger equation for $|\Psi\rangle$ then reads
\begin{align}
  \frac{\mathrm{d}}{\mathrm{d}t} u_a =& -\mathrm{i}g u_b - \mathrm{i}\sqrt{\kappa_a}\alpha_t
  \label{eq_modeaEvolEq}\\
  \frac{\mathrm{d}}{\mathrm{d}t} u_b =& -\mathrm{i}g u_a - \mathrm{i}\sqrt{\kappa_b}\beta_t
  \label{eq_modebEvolEq}\\
  \frac{\mathrm{d}}{\mathrm{d}t} \alpha_{\tau} =& -\mathrm{i}\sqrt{\kappa_a}c_a\delta(t-\tau)\label{eq_fieldaEvolEq}\\
  \frac{\mathrm{d}}{\mathrm{d}t} \beta_{\tau} =& -\mathrm{i}\sqrt{\kappa_a}c_b\delta(t-\tau).\label{eq_fieldbEvolEq}
\end{align}

We define the input and output wave packets as 
\begin{align*}
  \alpha_{\mathrm{in}}(\tau) =& \lim_{t\to -\infty}\alpha_{\tau}\\
  \alpha_{\mathrm{out}}(\tau) =& \lim_{t\to +\infty}\alpha_{\tau}\\
  \beta_{\mathrm{in}}(\tau) =& \lim_{t\to -\infty}\beta_{\tau}\\
  \beta_{\mathrm{out}}(\tau) =& \lim_{t\to +\infty}\beta_{\tau}.
\end{align*}
Integration of equations (\ref{eq_fieldaEvolEq}) and (\ref{eq_fieldbEvolEq}) yields the input-output relations
\begin{align*}
  \alpha_{\mathrm{out}}(\tau) =& \alpha_{\mathrm{in}}(\tau) - \mathrm{i}\sqrt{\kappa_a} c_a(\tau)\\
  \beta_{\mathrm{out}}(\tau) =& \beta_{\mathrm{in}}(\tau) - \mathrm{i}\sqrt{\kappa_b} c_b(\tau),
\end{align*}
as well as $\alpha_{t}(t) = \alpha_{\mathrm{in}}(t) - \mathrm{i}\sqrt{\kappa_a} c_a(t)/2$ and analogously for $\beta_{t}(t)$. Substituting these relations into evolution equations (\ref{eq_modeaEvolEq}) and (\ref{eq_modebEvolEq}) allows us to write a closed set of equations for only $u_a$ and $u_b$ in terms of the input wave packet functions:
\begin{align}
  \frac{\mathrm{d}}{\mathrm{d}t} u_a =& -\mathrm{i}g u_b - \frac{1}{2}\kappa_a u_a
  - \mathrm{i}\sqrt{\kappa_a}\alpha_{\mathrm{in}}
  \label{eq_modeaEvolEq2}\\
  \frac{\mathrm{d}}{\mathrm{d}t} u_b =& -\mathrm{i}g u_a - \frac{1}{2}\kappa_b u_b
  - \mathrm{i}\sqrt{\kappa_b}\beta_{\mathrm{in}}
  \label{eq_modebEvolEq2}
\end{align}

We will solve these equations in frequency space under the assumption of $C=4g^2/\kappa_a\kappa_b = 1$. This results in
\begin{align}
  u_a(\omega) =& \sqrt{\kappa_a}\frac{
  \left(\omega+\frac{\mathrm{i}\kappa_b}{2}\right)\alpha_{\mathrm{in}}(\omega) + \frac{1}{2}\kappa_b\beta_{\mathrm{in}}(\omega)
  }{\left(\omega+\frac{\mathrm{i}\kappa_a}{2}\right)\left(\omega+\frac{\mathrm{i}\kappa_b}{2}\right) - \frac{\kappa_a\kappa_b}{4}}
  \label{eq_freqSpaceEqn_ua}\\
  u_b(\omega) =& \sqrt{\kappa_b}\frac{
  \left(\omega+\frac{\mathrm{i}\kappa_a}{2}\right)\beta_{\mathrm{in}}(\omega) + \frac{1}{2}\kappa_a\alpha_{\mathrm{in}}(\omega)
  }{\left(\omega+\frac{\mathrm{i}\kappa_a}{2}\right)\left(\omega+\frac{\mathrm{i}\kappa_b}{2}\right) - \frac{\kappa_a\kappa_b}{4}}
  \label{eq_freqSpaceEqn_ub}\\
  \alpha_{\mathrm{out}}(\omega) =& \frac{
  \omega\left(\omega-\frac{\mathrm{i}(\kappa_a-\kappa_b)}{2}\right)\alpha_{\mathrm{in}}(\omega) - 
  \frac{\mathrm{i}\kappa_a\kappa_b}{2}\beta_{\mathrm{in}}(\omega)
  }{\left(\omega+\frac{\mathrm{i}\kappa_a}{2}\right)\left(\omega+\frac{\mathrm{i}\kappa_b}{2}\right) - \frac{\kappa_a\kappa_b}{4}}\\
  \beta_{\mathrm{out}}(\omega) =& \frac{
  \omega\left(\omega-\frac{\mathrm{i}(\kappa_b-\kappa_a)}{2}\right)\beta_{\mathrm{in}}(\omega) - 
  \frac{\mathrm{i}\kappa_a\kappa_b}{2}\alpha_{\mathrm{in}}(\omega)
  }{\left(\omega+\frac{\mathrm{i}\kappa_a}{2}\right)\left(\omega+\frac{\mathrm{i}\kappa_b}{2}\right) - \frac{\kappa_a\kappa_b}{4}}.
\end{align}

For signals on resonance ($\omega=0$) with a small bandwidth this results in principle in perfect conversion with $\alpha_{\mathrm{out}} =\mathrm{i}\beta_{\mathrm{in}}$ and $\beta_{\mathrm{out}} = \mathrm{i}\alpha_{\mathrm{in}}$.

The full-width-half maximum bandwidth $\Delta$ of the converter defined by $|\partial\alpha_{\mathrm{out}}/\partial\alpha_{\mathrm{in}}|_{\omega=\pm\Delta/2}^2 = 1/2$ is given by
\begin{align}
  \Delta^2 &= 2\left(
  \sqrt{\kappa_a^2\kappa_b^2 + \left(\frac{\kappa_a-\kappa_b}{2}\right)^4} -
  \left(\frac{\kappa_a-\kappa_b}{2}\right)^2
  \right)\label{eq_conversionBW}\\
  &\approx 2\kappa_a\text{ if }\kappa_a\ll\kappa_b\nonumber\\
  &\approx 2\kappa_b\text{ if }\kappa_b\ll\kappa_a\nonumber
\end{align}

\section{The full RWA Hamiltonian}\label{app_fullHamiltonian}
For reference, we show the full form of the $\op{\phi}^4$ term in the Hamiltonian after performing the rotating wave approximation.

\begin{align*}
 \op{H} =& \op{H}_0 - \frac{1}{4L^3 I_{*}^2} :\left[
 6\phi_a^4 \op{a}^{\dagger} \op{a}^{\dagger} \op{a} \op{a} +
 6\phi_b^4 \op{b}^{\dagger} \op{b}^{\dagger} \op{b} \op{b} +
 6\phi_c^4 \op{c}^{\dagger} \op{c}^{\dagger} \op{c} \op{c} \right.\\
 &
 + 24\phi_a^2\phi_b^2 \op{a}^{\dagger}\op{a} \op{b}^{\dagger} \op{b} +
 24\phi_a^2\phi_c^2 \op{a}^{\dagger}\op{a} \op{c}^{\dagger} \op{c} +
 24\phi_b^2\phi_c^2 \op{b}^{\dagger}\op{b} \op{c}^{\dagger} \op{c}\\
 & \left.
 + 12\phi_a\phi_b\phi_c^2 (\op{a} \op{b}^{\dagger} \op{c} \op{c} + 
 \op{a}^{\dagger} \op{b} \op{c}^{\dagger} \op{c}^{\dagger})
 \right]:\\
 =& \op{H}_0 - \frac{1}{4L^3 I_{*}^2} \left(
 6\phi_a^4 \op{a}^{\dagger} \op{a}^{\dagger} \op{a} \op{a} +
 6\phi_b^4 \op{b}^{\dagger} \op{b}^{\dagger} \op{b} \op{b} +
 6\phi_c^4 \op{c}^{\dagger} \op{c}^{\dagger} \op{c} \op{c} \right.\\
 &
 + 24\phi_a^2\phi_b^2 \op{a}^{\dagger}\op{a} \op{b}^{\dagger} \op{b} +
 24\phi_a^2\phi_c^2 \op{a}^{\dagger}\op{a} \op{c}^{\dagger} \op{c} +
 24\phi_b^2\phi_c^2 \op{b}^{\dagger}\op{b} \op{c}^{\dagger} \op{c}\\
 & 
 + 12\phi_a\phi_b\phi_c^2 (\op{a} \op{b}^{\dagger} \op{c} \op{c} + 
 \op{a}^{\dagger} \op{b} \op{c}^{\dagger} \op{c}^{\dagger})\\
 &
 + 12(\phi_a^2+\phi_b^2+\phi_c^2)(\phi_a^2 \op{a}^{\dagger} \op{a} + 
 \phi_b^2 \op{b}^{\dagger} \op{b} +\phi_c^2 \op{c}^{\dagger} \op{c})\\
 & + \left.
 3(\phi_a^2+\phi_b^2+\phi_c^2)^2
 \right).
\end{align*}
Here $:[\ldots]:$ denotes symmetric ordering of the ladder operators. Transforming all the terms into normal order results in the second expression. Terms quadratic in the ladder operators can be absorbed into $\op{H}_0$ and the remaining higher order terms written in an abbreviated form as
\begin{align*}
  \op{H} =& \op{H}_0 + 
  \frac{1}{2}
  \hspace{-1mm}\sum_{A \in \{a,b,c\}}\hspace{-2mm} 
  \chi_A \op{A}^{\dagger}\op{A}^{\dagger}\op{A}\op{A} +
  \hspace{-5mm}\sum_{A<B \in\{a,b,c\}}\hspace{-5mm} 
  \chi_{AB} \op{A}^{\dagger}\op{A} \op{B}^{\dagger}\op{B}\\
  & + g_0(\op{a}\op{b}^{\dagger}\op{c}\op{c} + 
  \op{a}^{\dagger}\op{b} \op{c}^{\dagger}\op{c}^{\dagger})\text{, where}\\
  \chi_A =& -\frac{3\phi_{A}^4}{L^3 I_{*}^2},\\
  \chi_{AB} =& -\frac{6\phi_{A}^2\phi_{B}^2}{L^3 I_{*}^2},\\
  g_0 =& -\frac{3\phi_a\phi_b\phi_c^2}{L^3 I_{*}^2}.
\end{align*}

After replacing $\op{c}$ by $\op{c}+\sqrt{\nPump}$ with $\nPump$ chosen such that terms linear in $\op{c}$ and $\op{c}^{\dagger}$ exactly cancel the terms in $\op{H}_0$ due to the coherent pump, and after absorbing terms proportional to $\op{a}^{\dagger}\op{a}$, $\op{b}^{\dagger}\op{b}$ and $\op{c}^{\dagger}\op{c}$ into $\op{H}_0$, we get
\begin{align*}
  \op{H} =& \op{H}_0 + 
  \frac{1}{2}\chi_a \op{a}^{\dagger}\op{a}^{\dagger}\op{a}\op{a} +
  \frac{1}{2}\chi_b \op{b}^{\dagger}\op{b}^{\dagger}\op{b}\op{b} +
  \chi_{ab} \op{a}^{\dagger}\op{a} \op{b}^{\dagger}\op{b}\\
  & + \nPump g_0(\op{a}\op{b}^{\dagger}+\op{a}^{\dagger}\op{b}) + 
  \nPump \chi_c(\op{c}^{\dagger}\op{c}^{\dagger}+\op{c}\op{c})\\
  & + \sqrt{\nPump}[(\chi_{ac}\op{a}^{\dagger}\op{a} + 
  \chi_{bc}\op{b}^{\dagger}\op{b} + g \op{a}^{\dagger}\op{b})\op{c}^{\dagger} 
    + \mathrm{H.c.}]\\
  & + 2\sqrt{\nPump}\chi_c (\op{c}^{\dagger}\op{c}^{\dagger}\op{c} + \mathrm{H.c.})\\
  & + \frac{1}{2}\chi_c \op{c}^{\dagger}\op{c}^{\dagger}\op{c}\op{c}\\
  & + \chi_{ac}\op{a}^{\dagger}\op{a} \op{c}^{\dagger}\op{c} + 
  \chi_{bc}\op{b}^{\dagger}\op{b} \op{c}^{\dagger}\op{c}.
\end{align*}

If $\nPump\chi_c\ll \kappaPump$, mode $c$ can be approximated as a linear system in its (displaced) ground state and the squeezing terms $\op{c}^{\dagger}\op{c}^{\dagger}+\op{c}\op{c}$ as well as $\op{c}^{\dagger}\op{c}^{\dagger}\op{c}+\op{c}^{\dagger}\op{c}\op{c}$ and $\op{c}^{\dagger}\op{c}^{\dagger}\op{c}\op{c}$ may be neglected. From the terms which couple $a$ and $b$ to $c$, we only keep the ones of highest order in $\nPump$. Moreover, we will assume that at most a single photon exists in modes $a$ and $b$ at any time. This also allows us to neglect the Kerr terms $\op{a}^{\dagger}\op{a}^{\dagger}\op{a}\op{a}$, $\op{b}^{\dagger}\op{b}^{\dagger}\op{b}\op{b}$ and $\op{a}^{\dagger}\op{a} \op{b}^{\dagger}\op{b}$. We are then left with
\begin{align}\label{eq_converterHam}
  \op{H} =& \op{H}_0 + \nPump g_0(\op{a} \op{b}^{\dagger}+\op{a}^{\dagger}\op{b})\\
  & + \sqrt{\nPump}[(\chi_{ac}\op{a}^{\dagger}\op{a} + 
  \chi_{bc}\op{b}^{\dagger}\op{b} + g_0 \op{a}^{\dagger}\op{b})\op{c}^{\dagger} 
    + \mathrm{H.c.}].\nonumber
\end{align}

The terms on the first line describe the ideal operation of the converter while the second line represents in principle unwanted coupling between the converter modes $a$ and $b$ and quantum fluctuations of the pump mode $c$.

\section{Derivation of converter dephasing due to pump shot noise}\label{app_measDeph}

To estimate how the shot noise of the pump tone affects the performance of the converter, we need to take into account some of the terms from the $\op{\phi}^4$ nonlinearity which we have neglected when analyzing the ideal operation of the device. The largest among them is the coupling term proportional to $\sqrt{\nPump}$ shown on the second line of Eq.~(\ref{eq_converterHam}). This term describes back-action of the converted photons on the pump field. This results in some degree of entanglement between the two and therefore dephasing of the converter's reduced density matrix. 

To describe this effect, we first note that the characteristic relaxation timescale of mode $c$, determined by its linewidth $\kappaPump$, will typically be much faster than the evolution of modes $a$ and $b$. This means that we can approximate mode $c$ as a Markovian bath and write an effective master equation for $a$ and $b$ and their input/output modes. We will neglect the weak nonlinearity of mode $c$ and approximate its steady state by the vacuum in the displaced reference frame. We will further assume that the dynamics of modes $a$ and $b$ can be neglected on time-scales over which the correlation functions of $c$ relax to their steady state values. We can then perform the standard master equation derivation with a system-bath coupling given by $\op{K} \op{c}^{\dagger} + \op{K}^{\dagger} \op{c}$, where
\[
  \op{K} = \sqrt{\nPump}(\chi_{ac}\op{a}^{\dagger}\op{a} + 
  \chi_{bc}\op{b}^{\dagger}\op{b} + g_0 \op{a}^{\dagger}\op{b}).
\]
This results in
\begin{align*}
  \frac{\mathrm{d}}{\mathrm{d}t}
  \op{\rho}_{ab} =& -\mathrm{i}[H_0+g(\op{a}\op{b}^{\dagger}+\op{a}^{\dagger}\op{b}),
  \op{\rho}_{ab}]\\
  & 
  - \op{K}^{\dagger} \op{K}\op{\rho}_{ab}\frac{1}{\kappaPump/2+\mathrm{i}\delta} 
  - \op{\rho}_{ab} \op{K}^{\dagger} \op{K}\frac{1}{\kappaPump/2-\mathrm{i}\delta} \\
  &
  + \op{K}\op{\rho}_{ab} \op{K}^{\dagger} \left(
  \frac{1}{\kappaPump/2+\mathrm{i}\delta} +
  \frac{1}{\kappaPump/2-\mathrm{i}\delta} 
  \right)\\&\\
  =& -\mathrm{i}[\op{H}_0+g(\op{a}\op{b}^{\dagger}+\op{a}^{\dagger}\op{b})+
  \Delta \op{H},\op{\rho}_{ab}]\\
  &+ \mathcal{D}_{q\op{K}}[\op{\rho}_{ab}]\text{, where}\\&\\
  q =& \sqrt{\frac{\kappaPump}{\delta^2+\kappaPump^2/4}},\\
  \Delta \op{H} =& -\frac{\delta}{\delta^2+\kappaPump^2/4}\op{K}^{\dagger}\op{K}.
\end{align*}
Here $\delta$ is the detuning of mode $c$ in the pump reference frame.

Instead of solving this master equation for the full density matrix $\op{\rho}_{ab}$, we will consider the purity $\mathrm{Tr}\,\op{\rho}_{ab}^2$  as a useful figure of merit quantifying the decoherence of the converted photons. It evolves according to
\begin{align*}
  \frac{\mathrm{d}}{\mathrm{d}t} \mathrm{Tr}\,\op{\rho}_{ab}^2 =&
  2\mathrm{Tr}\,\frac{\mathrm{d}\op{\rho}_{ab}}{\mathrm{d}t}\op{\rho}_{ab} \\
  =& 2q^2\mathrm{Tr}(\op{K}\op{\rho}_{ab}\op{K}^{\dagger}\op{\rho}_{ab} -
  \op{K}^{\dagger}\op{K}\op{\rho}_{ab}\op{\rho}_{ab}).
\end{align*}

We will calculate the purity to lowest order in $\op{K}$ by integrating this equation and approximating $\op{\rho}_{ab}$ on its right-hand side by the pure state $|\Psi\rangle$ unitarily evolving under the Hamiltonian $\op{H}_0 + g(\op{a}\op{b}^{\dagger}+\op{a}^{\dagger}\op{b})$. The final state purity at $t\to +\infty$ is then
\begin{equation}\label{eq_purity}
  \mathrm{Tr}\,\op{\rho}_{ab}^2 = 1 - 2q^2\int_{-\infty}^{+\infty}
  \langle \op{K}^{\dagger}\op{K} \rangle - \langle \op{K}^{\dagger}\rangle \langle \op{K}\rangle
  \,\mathrm{d}t.
\end{equation}

As derived in appendix \ref{app_convEvolution}, the state of the system has the form
\[
  |\Psi\rangle = \left(u_a \op{a}^{\dagger} + u_b \op{b}^{\dagger} + \ldots\right)|0\rangle,
\]
where the omitted terms contain photons in the input/output fields which do not contribute to the expectation values of $\op{K}$ and $\op{K}^{\dagger}K$. The evolution of the coefficients $u_a$ and $u_b$ is given in frequency space by equations (\ref{eq_freqSpaceEqn_ua}) and (\ref{eq_freqSpaceEqn_ub}). We will assume that the converted photon has a very narrow bandwidth. Under this assumption, the frequency dependence of the converter's response can be neglected and we have
\begin{align*}
  u_a(t) = & -(\beta_{\mathrm{in}}(t)+\mathrm{i}\alpha_{\mathrm{in}}(t)) / \sqrt{\kappa_a}\\
  u_b(t) = & -(\alpha_{\mathrm{in}}(t)+\mathrm{i}\beta_{\mathrm{in}}(t)) / \sqrt{\kappa_b}.
\end{align*}

Note that this means that each of the expectation values in the expression for the purity is a quadratic function of the the input wave packet functions. Due to the normalization condition $\int|\alpha_{\mathrm{in}}|^2\,\mathrm{d}t + \int|\beta_{\mathrm{in}}|^2\,\mathrm{d}t=1$, the two functions scale as $1/\sqrt{T}$ with the characteristic time scale $T$ of the wave packets. This implies that $\langle \op{K}^{\dagger}\op{K}\rangle\sim 1/T$ and $\langle \op{K}^{\dagger}\rangle\langle \op{K}\rangle\sim 1/T^2$, so we can neglect the term $\langle \op{K}^{\dagger}\rangle\langle \op{K}\rangle$ in the limit of narrow-band wave packets.

By substituting 
\begin{align*}
  \langle \op{K}^{\dagger}\op{K}\rangle =&
  \nPump\chi_{ac}^2 |u_a|^2 + \nPump(\chi_{bc}^2+g_0^2) |u_b|^2\\
  & + \nPump g_0\chi_{ac}(u_a^{*}u_b + u_b^{*} u_a)
\end{align*}
into Eq.~(\ref{eq_purity}), we get
\begin{equation}\label{eq_measInducedDeph}
  \mathrm{Tr}\,\op{\rho}_{ab}^2 = 1 - \frac{2\nPump\kappaPump}{\delta^2 + \kappaPump^2/4} \left(
  \frac{\chi_{ac}^2}{\kappa_a} + \frac{\chi_{bc}^2+g_0^2}{\kappa_b}
  \right).
\end{equation}
Here we made the assumption that either $\alpha_{\mathrm{in}}(t)=0$ or $\beta_{\mathrm{in}}(t)=0$ for all $t$. In other words, that we are sending a photon either from the microwave or mm wave side but not both at the same time. The terms in this expression can be understood intuitively as follows. The decoherence is due to the pump field carrying away information about the state of the converter modes $a$ and $b$. A leaky resonator coupled to a quantum system causes a measurement-induced dephasing at a rate given by $\kappaPump|\delta c|^2/2$, where $\delta c$ is the difference in the displacements of the pointer states caused by a change in the quantum system's state. In our case, addition of a photon into mode $a$ results in a frequency shift $\chi_{ac}$ of mode $c$ which leads to a displacement shift of $|\delta c|=\chi_{ac}\sqrt{\nPump}/\sqrt{\delta^2+\kappaPump^2/4}$. The converted photon spends on average a time on the order of $1/\kappa_a$ in mode $a$, accumulating a dephasing proportional to $\nPump\kappaPump\chi_{ac}^2/\kappa_a(\delta^2+\kappaPump^2/4)$ due to measurement of mode $a$ by the pump. This is exactly the first term in the purity loss expression on the right-hand side of our equation above. Similarly, the second term in the brackets results from measurement-induced dephasing in mode $b$. The additional term with $g_0^2$ is due to the third type of dephasing process where a photon can be swapped from $b$ to $a$ while two additional photons are created in the pump field.

\section{Back-action noise heating}

In addition to the terms considered above that are energy conserving and cause dephasing and scale as $\sqrt{\nPump}$, there are terms which were neglected as they are approximately non-energy conserving (by $\approx \omegaMW$), but may still be significant since they scale as $\nPump$. We consider the spontaneous emission process which may only be partially filtered by the cavity and becomes significant at larger $\kappaMM/\omegaMW$~\cite{Chu1985,Wilson-Rae2007,Marquardt2007}. In the full RWA Hamiltonian we neglected the term $\op{H}_\textrm{noise} = \exp(2i \omegaMW t)g\op{b}^\dagger\op{a}^\dagger  + \textrm{h.c.}$. Consider a highly simplified conversion Hamiltonian $\op{H}=g(\opd{a}\op{b}+\opd{b}\op{a})+\op{H}_\textrm{noise}$, which can be written in a time-independent way as
\begin{equation}
\op{H}=\omegaMW (\opd{a}\op{a} + \opd{b}\op{b}) + g(\opd{a}+\op{a})(\opd{b}+\op{b}).
\end{equation}
By solving the Heisenberg-Langevin equations, we find that the spectral density of the microwave mode with vacuum bath inputs at the matching condition $4g^2=\kappaMM\kappaMW$  in the limit where $\kappaMM\gg\kappaMW,g$ is given by~\cite{Safavi-Naeini2013a}
\begin{eqnarray}
S_{\op{a}\op{a}}(\omega)&=\int_{-\infty}^\infty d\omega^\prime \langle \opd{a}(\omega)\op{a}(\omega^\prime)\rangle \\
&= \frac{2\kappaMW}{(\omega+\omegaMW)^2+(\kappaMW)^2} n_\text{heating}.
\end{eqnarray}
with $n_\text{heating} = \frac{1}{2}\left(\frac{\kappaMM}{4\omegaMW}\right)^2$. The converter's microwave mode temperature with vacuum inputs approaches a steady-state photon occupation of $\langle \opd{a}\op{a}\rangle = n_\text{heating}$ due to the emission into the lower frequency sideband of the mm-wave resonator. This means that microwave output for a vacuum input will be thermal radiation at this temperature. This allows us to calculate the purity of the converted state for vacuum input. For very small heating rates, $\op{\rho}_\text{out} = (1-n_\text{heating})|0\rangle\langle0| + n_\text{heating}|1\rangle\langle1|$ so $\textrm{Tr}\,\op{\rho}_\text{out} \opd{a}\op{a} = n_\text{heating}$. The purity of this state is
\begin{equation}\label{eq_heatingPurityLoss}
  \textrm{Tr}\,\op{\rho}_\text{out}^2 \approx1-\left(\frac{\kappaMM}{4\omegaMW}\right)^2.
\end{equation}

\section{Perturbative expansion of a nonlinear resonator's steady state}\label{app_NLcorrections}
To find how the steady state of a driven lossy nonlinear resonator deviates from a coherent state with increasing mean photon number, we perform a perturbative expansion in the Kerr nonlinearity $\chi_c$. We write the Liouvillian of the system as $\mathcal{L}=\mathcal{L}_0+\delta\mathcal{L}$, where $\mathcal{L}_0$ is the generator of a linear resonator's dissipative evolution and $\delta\mathcal{L}[\op{\rho}]=-\mathrm{i}\chi_c[\op{c}^{\dagger}\op{c}^{\dagger}\op{c}\op{c},\op{\rho}]/2$. If we expand the density matrix $\op{\rho}$ in powers of $\chi_c$ as $\op{\rho}=\op{\rho}_0+\op{\rho}_1+\ldots$, and require the steady-state equation $\mathcal{L}[\op{\rho}]=0$ to be satisfied to all orders, we get the following recursive relations for $\op{\rho}_j$:
\begin{align}
  \mathcal{L}_0[\op{\rho}_0] &= 0,\\
  \mathcal{L}_0[\op{\rho}_{j+1}] &= -\delta\mathcal{L}[\op{\rho}_j]\text{ for }j>0.\label{eq_steadyStatePertEq}
\end{align}
$\op{\rho}_0$ is therefore the steady state of a linear resonator -- a coherent state $|\alpha_0\rangle\langle\alpha_0|$. The higher order corrections can then be found one by one by successively solving Eq.~(\ref{eq_steadyStatePertEq}). This equation does not have a unique solution because the superoperator $\mathcal{L}_0$ has a nontrivial null space spanned by $\op{\rho}_0$. We can, however, find $\op{\rho}_{j+1}$ uniquely by further requiring that $\mathrm{Tr}\,\op{\rho}=1$ to all orders, that is, $\mathrm{Tr}\,\op{\rho}_j=0$ for $j>0$.

To find the solution of Eq.~(\ref{eq_steadyStatePertEq}), we note that if we define displaced ladder operators $\op{C}=\op{c}-\alpha_0$, $\mathcal{L}_0$ preserves each of the subspaces $\mathcal{S}_{m}$ (for $m=0,\pm 1,\pm 2,\ldots$) spanned by the basis $B_m = \{(\op{C}^{\dagger})^{k}\op{\rho}_0 \op{C}^l : k-l = m\}$. In fact, it maps each $(\op{C}^{\dagger})^{k}\op{\rho}_0 \op{C}^l$ with the exception of $k=l=0$ onto a linear combination of itself and (if $k,l>0$) $(\op{C}^{\dagger})^{k-1}\op{\rho}_0 \op{C}^{l-1}$. In other words, the matrix corresponding to $\mathcal{L}_0$ in the basis $B_m$ is upper triangular (with only a single off-diagonal band) and has a  zero on the first diagonal position for $m=0$. This means that we can always find a solution $\op{\sigma}$ such that $\mathcal{L}_0[\op{\sigma}]$ matches any given right-hand side up to some multiple of $\op{\rho}_0$. The coefficient of this multiple is fixed by the property $\mathrm{Tr}\,\mathcal{L}_0=0$ and therefore a solution exists for any traceless right-hand side. Since the expression on the right-hand side of Eq.~(\ref{eq_steadyStatePertEq}) is a commutator, this requirement is satisfied identically.

The calculation of the individual corrections $\op{\rho}_j$ is then rather straightforward but leads to very long expressions. We have used the SymPy package for Python~\cite{Meurer2017} to calculate the first and second-order corrections $\op{\rho}_1$ and $\op{\rho}_2$. We could then evaluate the overlap $\langle\alpha|\op{\rho}|\alpha\rangle$ as a function of $\alpha$ and $\chi_c$ to second order in $\chi_c$. Since the overlap has a global maximum for $\alpha=\alpha_0$ and $\chi_c=0$, we expanded it around this point into $\langle\alpha|\op{\rho}|\alpha\rangle = 1 - K(\alpha-\alpha_0,\chi_c)$, where $K$ is quadratic in $\alpha-\alpha_0$ and $\chi_c$. Finally, we found the minimum of $K$ with respect to $\alpha-\alpha_0$ for fixed $\chi_c$ which then yielded an approximation for the maximum overlap
\[
  \max_{\alpha}\langle\alpha|\op{\rho}|\alpha\rangle = 
  1 - \frac{3\nPump^2\chi_c^2}{2(4\delta^2+\kappaPump^2)}
\]

\bibliography{LINQS_mmWave_quantum_conversion}

\end{document}